\documentclass[reqno]{amsart}
\usepackage{amssymb, latexsym, euscript}
\usepackage[cp1251]{inputenc}
\usepackage[english,ukrainian]{babel}
\newtheorem{theorem}{Theorem}[section]
\newtheorem{proposition}[theorem]{Proposition}

\newtheorem{lemma}[theorem]{Lemma}
\newtheorem{definition}[theorem]{Definition}
\newtheorem{corollary}[theorem]{Corollary}

\theoremstyle{definition}

\newtheorem{remark}[theorem]{Remark}

\numberwithin{equation}{section}

\begin{document}

\title{On $J$-self-adjoint extensions of the Phillips symmetric operator}

\author[S.~Kuzhel]{S.~Kuzhel}
\author[O.~Shapovalova]{O.~Shapovalova}
\author[L.~Vavrykovych]{L.~Vavrykovych}

\address{Institute of Mathematics of the National
Academy of Sciences of Ukraine, 3 Tereshchenkivska Street, 01601,
Kiev-4 Ukraine} \email{kuzhel@imath.kiev.ua}

\address{National Pedagogical Dragomanov University} \email{oks2074@mail.ru}

\address{Nizhin State University, 2 Kropyv'yanskogo Street, 16602 Nizhin,
Ukraine} \email{khvn@aport.ru}

\keywords{$J$-self-adjoint extensions, extension theory of symmetric operators, $C$-symmetry, $\mathcal{PT}$-symmetric quantum mechanics}
\subjclass[2000]{Primary 47A55, 47B25; Secondary 47A57,
81Q15}

\begin{abstract} $J$-self-adjoint extensions of the Phillips symmetric operator $S$ are
studied. The concepts of stable and unstable $C$-symmetry are introduced in the extension theory
framework. The main results are the following: if ${A}$ is a $J$-self-adjoint extension of $S$, then
either $\sigma({A})=\mathbb{R}$ or $\sigma({A})=\mathbb{C}$; if ${A}$ has a real spectrum, then ${A}$ has a stable $C$-symmetry
and ${A}$ is similar to a self-adjoint operator; there are no $J$-self-adjoint extensions of the Phillips operator with
unstable $C$-symmetry.
\end{abstract}

\maketitle
\section{Introduction} \label{sec1}
Let $\mathfrak{H}$ be a Hilbert space with inner product
$(\cdot,\cdot)$ and with fundamental symmetry $J$ (i.e., $J=J^*$ and
$J^2=I$). The space $\mathfrak{H}$ endowed with the indefinite inner product
(indefinite metric) \ $[x,y]_J:=(J{x}, y), \ \forall{x,y}\in\mathfrak{H}$ is
called  a Krein space  $(\mathfrak{H}, [\cdot,\cdot]_J)$.

An operator $A$ in ${\mathfrak H}$ is called $J$-{\it self-adjoint}
if $A$ is self-adjoint with respect to the indefinite metric
$[\cdot,\cdot]_J$. It is clear that $A$ is $J$-self-adjoint if and
only if
\begin{equation}\label{es1}
{A}^*J=J{A}.
\end{equation}

During the past ten years a steady interest in the study of
$J$-self-adjoint operators has been strongly increased by the
necessity of mathematically correct and rigorous analysis of
pseudo-Hermitian Hamiltonians arising in $\mathcal{PT}$-symmetric
quantum mechanics (PTQM) see e.g.
\cite{B4}-\cite{D2}, \cite{LT,MO,TT}.

In many cases, pseudo-Hermitian Hamiltonians admit the
representation $A+V$, where a (fixed) self-adjoint operator $A$ and
a non-symmetric potential $V$ satisfy certain (Krein space) symmetry
properties which allow one to formalize the expression $A+V$ as a
family of $J$-self-adjoint\footnote{under a special choice of
involution $J$} operators $A_\varepsilon$ acting in a Krein space
$({\mathfrak H}, [\cdot,\cdot]_J)$. Here
$\varepsilon\in\mathbb{C}^m$ is a complex parameter
characterizing the potential $V$.

Let $\Xi$ be the domain of variation of
$\varepsilon$. One of important problems for the collection
$\{A_\varepsilon\}$, which is directly inspirited by PTQM, is the
description of quantitative and qualitative changes of spectra
$\sigma(A_\varepsilon)$ when $\varepsilon$ runs $\Xi$. Nowadays this
topic has been analyzed with a wealth of technical tools (see, e.g.,
\cite{BKT,BKT1,GU1,GRS,Z}).

In particular, if the potential $V$ is singular, then operators
$A_\varepsilon$ turn out to be $J$-self-adjoint extensions of the
\emph{symmetric} operator $S=A\upharpoonright\ker{V}$
which \emph{commutes} with $J$ and spectral analysis of $A_\varepsilon$ can be carried out
by the extension theory methods \cite{A3,AK1,AK2,GorKoch}. Here,
the `main ingredients' are: a holomorphic operator function characterizing $S$
(the characteristic function $\Theta(\cdot)$ \cite{KO,KU1,SH} or the
Weyl function $M(\cdot)$ \cite{DHS,DHMS,DM}) and
 the boundary conditions which distinguish $A_\varepsilon$ among other $J$-self-adjoint extensions of $S$.
In such a setting, the spectral analysis of $A_\varepsilon$ is reduced to the routine solution of algebraic equations
including $\theta(\cdot)$ and boundary conditions.

In the present paper we are going to study a special case where
the characteristic function of a symmetric operator $S$ with finite deficiency indices is equal to zero ($\Theta(\mu)\equiv{0}$, \ $\forall{\mu}\in\mathbb{C}\setminus\mathbb{R}$).

One of general constructions leading to symmetric operators $S$ with the zero characteristic function is the following:
let $U$ be a bilateral shift with a wandering subspace $W_0$ in
${\mathfrak H}$ (see \cite{Sn1} for the terminology) and let $V$ be its restriction onto
$\mathfrak{H}\ominus{W_0}$, i.e.,
$V=U\upharpoonright(\mathfrak{H}\ominus{W_0})$. Then the operator
\begin{equation}\label{e1}
S=i(V+I)(V-I)^{-1}, \qquad
\mathcal{D}(S)=\mathcal{R}(V-I)
\end{equation}
is simple\footnote{An operator is called \emph{simple} if its restriction to any nontrivial reducing subspace is not a self-adjoint operator.} symmetric and its deficiency induces coincide with $<\dim{W_0},\dim{W_0}>$.

In other words, $S$ is the restriction of the Cayley
transform of $U$:
\begin{equation}\label{e2}
A=i(U+I)(U-I)^{-1}, \qquad \mathcal{D}(A)=\mathcal{R}(U-I)
\end{equation}
onto $\mathcal{D}(S)=\mathcal{R}(V-I)$.

The operator $S$ defined by (\ref{e1}),
(\ref{e2}) was used by Phillips \cite{PH} (with $\dim{W_0}=1$) as an
example of the symmetric operator, which is invariant with respect to
a certain set $\mathfrak{U}$ of unitary operators
($\mathfrak{U}$-invariant) but it has no $\mathfrak{U}$-invariant
self-adjoint extensions. For this reason, the simple symmetric operator $S$ determined by (\ref{e1}) and (\ref{e2})
will be referred as the \emph{Phillips symmetric operator}.

Due to specific properties of the Phillips operator (the characteristic function is zero, there are no real points of regular type of $S$, etc) we obtain an evolution of $\sigma(A_\varepsilon)$ which differs from the matrix models \cite{GU1,GRS,GSZ} and models based on $J$-self-adjoint
(symmetric) perturbations of the Schr\"{o}dinger or Dirac operator \cite{AMS,CGS,LT,Z}. For instance, in our case, either the spectrum of an
$J$-self-adjoint extension $A_\varepsilon$ of $S$ coincides with real line: $\sigma(A_\varepsilon)=\mathbb{R}$ or with complex plane: $\sigma(A_\varepsilon)=\mathbb{C}$ (Theorem \ref{tt1}).

One of the key points in PTQM is the description of a hidden
symmetry ${C}$  which exists for a given pseudo-Hermitian
Hamiltonian $A$ in the sector of exact $\mathcal{PT}$-symmetry
\cite{B6,B4,B1}. The operator $C$ has some rough analogy with the charge
conjugation operator in the quantum field theory \cite{B4} and it is determined non-uniquely \cite{BK1}.
The existence of ${C}$ gives rise to an inner product
$(\cdot,\cdot)_{{C}}=[{{C}}\cdot,\cdot]_J$ and the dynamics generated by $A$ is therefore
governed by a unitary time evolution.

For $J$-self-adjoint extensions $A_\varepsilon\supset{S}$, where $S$ is an \emph{arbitrary} symmetric operator commuting with $J$, we introduce the concepts of stable and unstable $C$-symmetry (Definition \ref{dad2}). These concepts are natural
in the extension theory framework. Roughly speaking, if $A_\varepsilon$ belongs to the sector $\Sigma_J^{\textsf{st}}$ of stable $C$-symmetry, then $A_\varepsilon$ preserves the property of $C$-symmetry under small variation of $\varepsilon$.

It follows from the results of \cite{AKG, GK1} that for some types of
singular perturbations of the Schr\"{o}dinger or the Dirac operator, the sector $\Sigma_J^{\textsf{unst}}$ of unstable $C$ symmetry
is not empty and operators $A_\varepsilon$ with real spectra and Jordan points correspond to the case
where $\varepsilon$ belongs to the boundary of $\Sigma_J^{\textsf{st}}$.

In the case of the Phillips symmetric operator $S$, the spectral picture above can be essentially simplified. Precisely, assuming the deficiency indices $<2,2>$ of $S$, we show that $\Sigma_J^{\textsf{unst}}=\varnothing$ and any $J$-self-adjoint extension of $S$ with real spectrum is similar to a self-adjoint operator (Theorem \ref{p14a} and Corollary \ref{uma1}).

Throughout the paper $\mathcal{D}(A)$, $\mathcal{R}(A)$, and
$\ker{A}$ denote the domain, the range, and the null-space of a
linear operator $A$, respectively, while
$A\upharpoonright{\mathcal{D}}$ stands for the restriction of $A$ to
the set $\mathcal{D}$. The set of points of regular type of a symmetric operator $S$ is
denoted by $\widehat{\rho}(S)$ (i.e.,
$r\in\widehat{\rho}(S)\iff\|(S-r{I})u\|\geq{k}\|u\|, \
\forall{u}\in\mathcal{D}(S), \ k>0$).

\setcounter{equation}{0}
\section{Preliminaries}
\label{sec2}
\subsection{Elements of the Krein space theory.}
Let $(\mathfrak{H}, [\cdot,\cdot]_J)$ be a Krein space with fundamental symmetry $J$.
The corresponding orthoprojectors  $P_{\pm}=\frac{1}{2}(I{\pm}J)$ determine \emph{the fundamental
decomposition} of $\mathfrak{H}$
\begin{equation}\label{d1}
\mathfrak{H}=\mathfrak{H}_+\oplus\mathfrak{H}_-, \qquad
\mathfrak{H}_-=P_{-}\mathfrak{H}, \quad \mathfrak{H}_+=P_{+}\mathfrak{H}.
\end{equation}

A subspace $\mathfrak{L}$ of $\mathfrak{H}$ is called \emph{hypermaximal neutral} if
$$
\mathfrak{L}=\mathfrak{L}^{[\bot]_J}=\{x\in\mathfrak{H}
 :  [x,y]_J=0, \ \forall{y}\in\mathfrak{L}\}.
 $$

A subspace $\mathfrak{L}\subset\mathfrak{H}$ is called {\it uniformly positive
(uniformly negative)} if $[x,x]_J\geq{a}^2\|x\|^2$ \
($-[x,x]_J\geq{a}^2\|x\|^2$) $a\in\mathbb{R}$ for all
$x\in\mathfrak{L}$. The subspaces $\mathfrak{H}_{\pm}$ in (\ref{d1}) are
examples of uniformly positive and uniformly negative subspaces and
they possess the property of maximality in the corresponding classes
(i.e., $\mathfrak{H}_{+}$ ($\mathfrak{H}_{-}$) does not belong as a subspace to any
uniformly positive (negative) subspace).

Let $\mathfrak{L}_+(\not={\mathfrak H}_+)$ be an arbitrary maximal uniformly positive subspace.
Then its $J$-orthogonal complement $ \mathfrak{L}_-=\mathfrak{L}_+^{[\bot]_J}$ is a
maximal uniformly negative and the direct $J$-orthogonal sum
\begin{equation}\label{d2}
\mathfrak{H}=\mathfrak{L}_+[\dot{+}]_J\mathfrak{L}_-
\end{equation}
gives another (then (\ref{d1})) decomposition of $\mathfrak{H}$ onto its
positive $\mathfrak{L}_+$ and negative $\mathfrak{L}_-$ parts (the brackets
$[\cdot]_J$ mean the orthogonality with respect to the indefinite
metric).

An arbitrary decomposition  of the Krein space $(\mathfrak{H},
[\cdot,\cdot]_J)$ onto its positive and negative parts (like
(\ref{d2})) is called \emph{canonical}.

The subspaces $\mathfrak{L}_\pm$ in (\ref{d2}) can be described as
$$
\mathfrak{L}_+=(I+X)\mathfrak{H}_+, \qquad \mathfrak{L}_-=(I+X^*)\mathfrak{H}_-,
$$
where $X:\mathfrak{H}_+\to\mathfrak{H}_-$ is a contraction and $X^*:\mathfrak{H}_-\to\mathfrak{H}_+$ is the adjoint of $X$.

The self-adjoint operator $T=XP_++X^*P_-$ acting in $\mathfrak{H}$ is called
{\it an operator of transition} from the fundamental decomposition
(\ref{d1}) to the canonical one (\ref{d2}). Obviously, $\mathfrak{L}_+=(I+T)\mathfrak{H}_+$
and $\mathfrak{L}_-=(I+T)\mathfrak{H}_-$.

Operators of transition admit a simple description. Namely, a
self-adjoint operator $T$ in  $\mathfrak{H}$ is an operator of transition if
and only if  $\|T\|<1$  and $JT=-TJ$.

The set $\{T\}$ of all possible operators of transition is in
one-to-one correspondence (via $\mathfrak{L}_\pm=(I+T)\mathfrak{H}_\pm$) with all possible
canonical  decompositions (\ref{d2}) of the Krein space $(\mathfrak{H},
[\cdot,\cdot]_J)$.

The projectors $P_{\mathfrak{L}_\pm} : \mathfrak{H}\to\mathfrak{L}_{\pm}$ onto $\mathfrak{L}_\pm$ with
respect to the decomposition (\ref{d2}) are determined by the
formulas
$$
P_{\mathfrak{L}_-}=(I-T)^{-1}(P_--TP_+), \quad
P_{\mathfrak{L}_+}=(I-T)^{-1}(P_+-TP_-).
$$

The bounded operator
\begin{equation}\label{eae2}
{C}=P_{\mathfrak{L}_+}-P_{\mathfrak{L}_-}=J(I-T)(I+T)^{-1}
\end{equation}
also describes subspaces ${\mathfrak{L}_\pm}$ in (\ref{d2}):
\begin{equation}\label{b4b}
\mathfrak{L}_+=\frac{1}{2}(I+{C})\mathfrak{H}, \qquad \mathfrak{L}_-=\frac{1}{2}(I-{C})\mathfrak{H}.
\end{equation}

The set of operators ${C}$ determined (\ref{eae2}) is completely
characterized by the conditions
\begin{equation}\label{sos1}
{C}^2=I, \qquad  J{C}>0.
\end{equation}

\subsection{Elements of the Von Neumann extension theory.}
Let $S$ be a closed symmetric densely defined
operator in a Hilbert space $\mathfrak{H}$ with equal (finite or
infinite) deficiency indices. Denote by
$\mathfrak{N}_i=\mathfrak{H}\ominus\mathcal{R}(S-iI)$
and
$\mathfrak{N}_{-i}=\mathfrak{H}\ominus\mathcal{R}(S+iI)$
the defect subspaces of $S$ and consider the Hilbert
space $\mathfrak{M}=\mathfrak{N}_{-i}\dot{+}\mathfrak{N}_{i}$ with
the inner product
$$
(f,g)_{\mathfrak{M}}=(f_i,g_i)+(f_{-i},g_{-i}) \quad
f=f_i+f_{-i}, \quad g=g_i+g_{-i} \quad \{f_{{\pm}i},
g_{{\pm}i}\}\subset\mathfrak{N}_{{\pm}i}.
$$

The operator $Z(f_{-i}+f_{i})=f_{-i}-f_{i}$ is a fundamental symmetry in the
Hilbert space $\mathfrak{M}$ and its restriction onto
$\mathfrak{N}_{-i}$ and $\mathfrak{N}_{i}$ coincide, respectively,
with $I$ and $-I$.

Let $J$ be a fundamental symmetry in $\mathfrak{H}$. In what follows we assume that
\begin{equation}\label{es12}
SJ=JS.
\end{equation}
Then the subspaces
$\mathfrak{N}_{\pm{i}}$ reduce $J$ and the restriction
$J\upharpoonright\mathfrak{M}$ gives rise to a fundamental symmetry in the
Hilbert space $\mathfrak{M}$. Moreover, according to the properties
of $Z$ mentioned above, $JZ=ZJ$. Therefore, $JZ$ is a fundamental symmetry in
$\mathfrak{M}$ and sesquilinear form
$$
[f,g]_{JZ}=(JZf,g)_{\mathfrak{M}}=(Jf_{-i},g_{-i})-(Jf_{i},g_{i})
$$
determines an indefinite metric on $\mathfrak{M}$.

According to von-Neumann formulas any closed intermediate extension
$A$ of $S$ (i.e.,
$S\subset{{A}}\subset{S}^*$) is
uniquely determined by the choice of a subspace
$M\subset\mathfrak{M}$. Precisely,
\begin{equation}\label{e55}
\mathcal{D}({A})=\mathcal{D}(S)\dot{+}M \qquad
\mbox{and} \qquad {A}=S^*\upharpoonright\mathcal{D}({A}).
\end{equation}
We use the notation $A_M$ for $J$-self-adjoint extensions
of $S$ determined by (\ref{e55}).

Let $A_{M}$ and $A_{\widetilde{M}}$ be arbitrary
extensions of $S$  that are defined by the subspaces
$M$ and $\widetilde{M}$, respectively. Taking (\ref{es12}) and
(\ref{e55}) into account we derive
\begin{equation}\label{e78a}
[A_{M}\psi, \phi]_{J}-[\psi, A_{\widetilde{M}}\phi]_{J}=2i[f,g]_{JZ}
\end{equation}
for all ${\psi=u+f}\in\mathcal{D}(A_{M}), \ f\in{M}, \quad
\phi=v+g\in\mathcal{D}(A_{\widetilde{M}}), \ g\in{\widetilde{M}}$.

It follows from (\ref{es1}) and (\ref{e78a}) that an extension $A_M$
of $S$ is $J$-self-adjoint if and only if
$$
 M={M}^{[\perp]_{JZ}}=\{f\in\mathfrak{M} \
: \ [f,g]_{JZ}=0, \ \forall{g}\in{M}\},
$$
i.e., if $M$ is a hypermaximal neutral subspace of the Krein space
$(\mathfrak{M}, [\cdot, \cdot]_{JZ})$. Formalizing this observation
we get the well-known result.
\begin{proposition}\label{t2}
The correspondence ${A}\leftrightarrow{M}$ determined by (\ref{e55})
is a bijection between $J$-self-adjoint (self-adjoint) extensions
$A$ of $S$ and hypermaximal neutral subspaces $M$ of the Krein space
$(\mathfrak{M}, [\cdot, \cdot]_{JZ})$ (of the Krein space $(\mathfrak{M}, [\cdot,
\cdot]_{Z})$).
\end{proposition}

Denote by $\Sigma_J(S)$ the set of $J$-self-adjoint
extensions of $S$. In general, these extensions may have complex
spectra and, moreover, the existence of ${A}\in\Sigma_J(S)$ with empty resolvent set (i.e.,
$\sigma({A})=\mathbb{C}$) is also possible. To guarantee nonempty resolvent set \emph{for any} ${A}\in\Sigma_J(S)$ we need
to impose additional constraints. In this way we recall that a $J$-self-adjoint operator $A$ is called
\emph{definitizable}  if the resolvent set of $A$ is nonempty and there exists a polynomial $p(\cdot)\not\equiv{0}$ such that
$p(A)$ is a nonnegative operator in the Krein space $({\mathfrak H}, [\cdot,\cdot]_J)$.

 \begin{proposition}\label{te2e}\cite{ABT}
Let $S$ have finite deficiency indices. Then if there exists a definitizable extension ${A}\in\Sigma_J(S)$, then an arbitrary operator from $\Sigma_J(S)$ has a nonempty resolvent set and is definitizable.
\end{proposition}

\subsection{Boundary value spaces technique.}
 Proposition \ref{t2} provides a description of $\Sigma_J(S)$ in terms of
the Krein space $(\mathfrak{M}, [\cdot, \cdot]_{JZ})$.
Another approach which allows one to avoid the
use of $\mathfrak{M}$ is based on the concept of
\emph{boundary triplets} (or boundary value spaces, see \cite{GorKoch} and the references therein).
 \begin{definition}\label{dd1}
A triplet $(\mathcal{H}, \Gamma_0, \Gamma_1)$, where $\mathcal{H}$
is an auxiliary Hilbert space and $\Gamma_0$, $\Gamma_1$ are linear
mappings of $\mathcal{D}(S^*)$ into $\mathcal{H}$,
is called a boundary triplet of $S^*$ if the abstract
Green identity
$$
(S^*\psi, \phi)-(\psi, S^*\phi)=(\Gamma_1\psi,
\Gamma_0\phi)_{\mathcal{H}}-(\Gamma_0\psi, \Gamma_1\phi)_{\mathcal{H}},
\quad  \psi, \phi\in\mathcal{D}(S^*)
$$
is satisfied and the map $(\Gamma_0,
\Gamma_1):\mathcal{D}(S^*)\to\mathcal{H}\oplus\mathcal{H}$
is surjective.
\end{definition}

 Denote
\begin{equation}\label{es15}
\mathfrak{N}_\mu=\mathfrak{H}\ominus\mathcal{R}(S-{\mu}I)=\ker(S^*-\overline{{\mu}}I), \quad \mu\in\widehat{\rho}(S).
\end{equation}

The Weyl function $M(\cdot)$ and the characteristic function
$\Theta(\cdot)$ of $S$ associated with a boundary triplet $(\mathcal{H}, \Gamma_0,
\Gamma_1)$ are defined as follows \cite{DM, Kochubei, SH}:
\begin{equation}\label{neww65}
\begin{array}{c}
M(\mu)\Gamma_0f_{\overline{\mu}}=\Gamma_1f_{\overline{\mu}} \quad
\forall{f}_{\overline{\mu}}\in\mathfrak{N}_{\overline{\mu}}, \quad
\forall\mu\in\widehat{\rho}(S), \vspace{4mm} \\
\Theta(\mu)(\Gamma_1+i\Gamma_0)f_{\overline{\mu}}=(\Gamma_1-i\Gamma_0)f_{\overline{\mu}},
\quad \forall\mu\in\widehat{\rho}(S) \ \ (\textsf{Im} \ \mu\geq0).
\end{array}
\end{equation}
 It is clear that $\Theta(\mu)=(M(\mu)-iI)(M(\mu)+iI)^{-1}, \
\forall\mu\in\widehat{\rho}(S) \ \ (\textsf{Im} \ \mu\geq0)$.

The Weyl function (or, characteristic function) determines a simple symmetric operator $S$ up
to unitary equivalence.

The simplest (\emph{canonical}) boundary triplet can immediately be
constructed as a triplet $(\mathfrak{N}_{-i}, \Gamma_0, \Gamma_1)$,
where
\begin{equation}\label{sas6}
\Gamma_0{\psi}=f_{-i}+Qf_{i}, \quad \Gamma_1{\psi}=if_{-i}-iQf_{i}, \quad
\psi=u+f_{-i}+f_{i}\in\mathcal{D}(S^*)
\end{equation}
and $Q$ is an arbitrary unitary mapping $Q :
\mathfrak{N}_{i}\to\mathfrak{N}_{-i}$.

To underline the dependence of $\Gamma_j$ on the choice of $Q$ in
(\ref{sas6}), we denote by $(\mathfrak{N}_{-i}, \Gamma_0,
\Gamma_1, Q)$ the corresponding boundary triplet.

If $Q$ commutes with $J$, then the boundary operators $\Gamma_j$ defined by (\ref{sas6}) satisfy the relations
\begin{equation}\label{ea6a}
\Gamma_0J=J\Gamma_0, \qquad \Gamma_1J=J\Gamma_1.
\end{equation}

By Proposition \ref{t2}, self-adjoint extensions
$A_M\supset{S}$ commuting with $J$ are described
by hypermaximal neutral subspaces
\begin{equation}\label{sas14}
M_G=\{ f_{i}+Gf_{i} \ | \ \forall{f_i}\in\mathfrak{N}_{i} \}
\end{equation}
of the Krein space $(\mathfrak{M}, [\cdot, \cdot]_{Z})$ which
satisfy the additional relation $JM_G=M_G$. Here $G
:\mathfrak{N}_{i}\to\mathfrak{N}_{-i}$ are unitary mappings.
Obviously, $JM_G=M_G\iff{JG=GJ}$. The latter gives rise to the
existence of boundary triplets $(\mathfrak{N}_{-i}, \Gamma_0,
\Gamma_1, -G)$ defined by (\ref{sas6}) with the additional properties
(\ref{ea6a}). We prove the following simple statement:
\begin{proposition}\label{cc2}
Boundary triplets $(\mathfrak{N}_{-i}, \Gamma_0, \Gamma_1, Q)$ satisfying
(\ref{ea6a}) exist if and only if the set of self-adjoint extensions
of $S$ commuting with $J$ is non-empty.
\end{proposition}

For such type of boundary triplets, Proposition \ref{t2} can be
rewritten as follows:
\begin{proposition}\label{tt34}
Let $(\mathfrak{N}_{-i}, \Gamma_0, \Gamma_1, Q)$ be a boundary triplet of $S^*$ which satisfies (\ref{ea6a}).
Then an arbitrary ${A}\in\Sigma_J(S)$ with $i\not\in\sigma(A)$ coincides with the restriction of
$S^*$ onto the domain
\begin{equation}\label{as1}
\mathcal{D}({A})=\{f\in\mathcal{D}(S^*) \ | \
K(\Gamma_1+i\Gamma_0)f=(\Gamma_1-i\Gamma_0)f\},
\end{equation}
where $K$ is a $J$-unitary operator in $\mathfrak{N}_{-i}$ (i.e.,
$J=K^*JK$).

The correspondence ${A}=A_K\leftrightarrow{K}$ determined by
(\ref{as1}) is a bijection between the set of all $J$-self-adjoint
extensions $A_K$ of $S$ such that
$i\not\in\sigma(A_K)$ and the set of $J$-unitary operators in
$\mathfrak{N}_{-i}$. Furthermore,
\begin{equation}\label{sos21}
A_K^*=A_{{(K^*)^{-1}}}.
\end{equation}
\end{proposition}

\begin{remark}\label{rr1} $J$-Self-adjoint extensions $A_M$ with
$i\in\sigma(A_M)$ are characterized by nontrivial intersections
$M\cap\mathfrak{N}_{-i}$ of the corresponding subspaces $M$ in
(\ref{e55}). In that case, the description (\ref{as1}) of
$\mathcal{D}(A_M)$ is impossible (since
$\ker(\Gamma_1-i\Gamma_0)=\mathfrak{N}_{-i}$ and
$\ker(\Gamma_1+i\Gamma_0)=\mathfrak{N}_{i}$ by (\ref{sas6})).
\end{remark}

\subsection{Description of $\Sigma_J(S)$. The case of deficiency indices $<2,2>$.}
We are going to analyze $\Sigma_J(S)$ in more detail for the case where $S$
has deficiency indices $<2,2>$. To avoid the study of self-adjoint extensions we assume $J\not=I$.
Then, the following subspaces of the Hilbert space $\mathfrak{M}$:
$$
\begin{array}{lr}
{\mathfrak{M}}_{++}=(I+Z)(I+J)\mathfrak{M}; \ & \ {\mathfrak{M}}_{--}=(I-Z)(I-J)\mathfrak{M}; \\
 {\mathfrak{M}}_{+-}=(I+Z)(I-J)\mathfrak{M}; \ & \
{\mathfrak{M}}_{-+}=(I-Z)(I+J)\mathfrak{M}
\end{array}
$$
are nontrivial and mutually orthogonal.  Therefore, $\dim{\mathfrak{M}}_{\pm\pm}=1$ (since $\dim{\mathfrak{M}}=4$) and
there exists an orthonormal basis $\{e_{\pm\pm}\}$ of the Hilbert space $\mathfrak{M}$ such
that
$$
{\mathfrak{M}}_{\pm\pm}=<e_{\pm\pm}>, \quad \mathfrak{N}_{-i}=<e_{++}, e_{+-}>, \quad \mathfrak{N}_{i}=<e_{-+}, e_{--}>.
$$
In that case
\begin{equation}\label{ess1}
\begin{array}{lr}
Je_{++}=e_{++}, \quad Je_{-+}=e_{-+}, & Je_{+-}=-e_{+-}, \quad Je_{--}=-e_{--};  \\
Ze_{++}=e_{++}, \quad Ze_{-+}=-e_{-+}, & Ze_{+-}=e_{+-}, \quad
Ze_{--}=-e_{--}.
\end{array}
\end{equation}

Let us consider the boundary triplet $(\mathfrak{N}_{-i}, \Gamma_0,
\Gamma_1, Q)$  defined by (\ref{sas6}), where a
unitary mapping $Q : \mathfrak{N}_{i}\to\mathfrak{N}_{-i}$ acts as follows:
\begin{equation}\label{sas92}
Qe_{-+}=e_{++}, \qquad Qe_{--}=e_{+-}.
\end{equation}
The operator $Q$ commutes with $J$ due to (\ref{ess1}) and hence,
relations (\ref{ea6a}) hold.

Denote by $\mathcal{K}=\|k_{ij}\|$ the matrix representation of a
$J$-unitary operator $K$ in $\mathfrak{N}_{-i}$ with respect to the
basis $\{e_{++},e_{+-}\}$. By (\ref{ess1}), the restriction of $J$
onto $\mathfrak{N}_{-i}$ can be identified (with respect to the
basis $\{e_{++},e_{+-}\}$) with the matrix  $\sigma_3=\left(\begin{array}{cc} 1  & 0 \\
0 & -1
\end{array}\right)$. This means that
$\sigma_3=\overline{\mathcal{K}^{t}}\sigma_3\mathcal{K}$ (since $K$
is $J$-unitary). The simple analysis of the latter relation leads to
the following description of $\mathcal{K}$:
\begin{equation}\label{as3a}
\mathcal{K}=\mathcal{K}(\zeta,
\phi, \omega, \xi)=e^{-i\xi}\left(\begin{array}{cc} -(\cosh\zeta)e^{-i\phi} & (\sinh\zeta)e^{-i\omega} \\
-(\sinh\zeta)e^{i\omega} & (\cosh\zeta)e^{i\phi} \end{array}\right),
\end{equation}
where $\zeta\in\mathbb{R}$ and  $\xi, \phi, \omega\in[0,2\pi)$.
Using Proposition \ref{tt34}, we obtain the following
\begin{proposition}\label{k12}
The formula
\begin{equation}\label{as1b}
S^*\upharpoonright\mathcal{D}(A_M), \qquad
\mathcal{D}(A_M)=\{f\in\mathcal{D}(S^*) \ | \
K(\Gamma_1+i\Gamma_0)f=(\Gamma_1-i\Gamma_0)f\},
\end{equation}
where $K$ is an arbitrary $J$-unitary operator in $\mathfrak{N}_{-i}$ and
the boundary triplet $(\mathfrak{N}_{-i}, \Gamma_0,
\Gamma_1, Q)$ is defined by (\ref{sas6}) and (\ref{sas92})
 establishes the one to one correspondence between
$J$-self-adjoint extensions $A_M\in\Sigma_J(S)$ with $i\not\in\sigma(A_M)$ and
matrices $\mathcal{K}(\zeta,
\phi, \omega, \xi)$ defined by (\ref{as3a}).
\end{proposition}

{\bf Remark.} It follows from Proposition \ref{t2} and relations
(\ref{ess1}) that operators $A_M\in\Sigma_J(S)$ with $i\in\sigma(A_M)$ are described by the two-parameter set of hypermaximal
neutral subspaces
$$
M(k_1,k_2)=<e_{++}+e^{ik_{1}}e_{+-}; \ e_{--}+e^{ik_{2}}e_{-+}>,
\qquad {k}_1, k_2\in\mathbb{R}
$$
of the Krein space $(\mathfrak{M}, [\cdot,\cdot]_{JZ})$. By
virtue of (\ref{sas6}) and (\ref{sas92}), the subspaces $M(k_1,k_2)$
can be (formally) described by (\ref{as1}) if we put
$$
\zeta=\infty, \quad
\xi=0, \quad \omega=\frac{k_2+k_1}{2}, \quad \phi=\frac{k_2-k_1}{2}
$$
in (\ref{as3a}) and consider $\cosh\infty=\sinh\infty=\infty$ as a
number.

To emphasize the relationship $A_M\leftrightarrow{{\mathcal{K}}}$ established in Proposition
\ref{k12}, we will use the notation $A_{\mathcal{K}}$ instead of $A_M$.

\begin{corollary}\label{p45}
The adjoint operator $A_{\mathcal{K}}^*$ of $A_{\mathcal{K}}\in\Sigma_J(S)$  is defined by $\mathcal{K}(-\zeta, \phi,
\xi, \omega)$ i.e.,
\begin{equation}\label{sdf1}
A_{\mathcal{K}(\zeta,
\phi, \omega, \xi)}^*=A_{\mathcal{K}(-\zeta,
\phi, \omega, \xi)}
\end{equation}

The set of self-adjoint extensions of $S$ commuting with $J$ is described
by unitary matrices ${\mathcal{K}(0,
\phi, \omega, \xi)}$.
\end{corollary}

{\it Proof.} The relation (\ref{sdf1}) follows from (\ref{sos21}) and (\ref{as3a}).

If a self-adjoint extension $A\supset{S}$ commutes
with $J$, then ${A}$ is also $J$-self-adjoint and
${A}\equiv{A}_{\mathcal{K}(\zeta, \phi, \omega, \xi)}$ by Proposition
\ref{k12}. Using (\ref{sdf1}) and taking into account (\ref{as3a}), we get $\zeta=0$  that completes the
proof of Corollary \ref{p45}.

\subsection{The property of $C$-symmetry.}
By analogy with \cite{B4} the definition of
${C}$-symmetry in the Krein spaces setting can be formalized as follows.
\begin{definition}\label{dad1}
An operator ${A}$ acting in a Krein space $(\mathfrak{H},
[\cdot,\cdot]_{{J}})$ has the property of ${C}$-symmetry if
there exists a bounded linear operator ${C}$ in $\mathfrak{H}$ such
that: \ $(i) \ {C}^2=I;$ \quad $(ii) \ {J}C>0$; \quad
$(iii) \ A{C}={C}A$.
\end{definition}

By virtue of (\ref{eae2}) and (\ref{sos1}) the property of
${C}$-symmetry of $A$ means that $A$ can be decomposed:
\begin{equation}\label{neww4}
A=A_+[\dot{+}]_{{J}}A_-, \qquad
A_+=A\upharpoonright_{\mathfrak{L}_+}, \quad
A_-=A\upharpoonright_{\mathfrak{L}_-}
\end{equation}
with respect to the canonical decomposition $(\ref{d2})$ (with subspaces
$\mathfrak{L}_\pm$ determined by (\ref{b4b})).

If a ${J}$-self-adjoint operator $A$ possesses the property
of $C$-symmetry, then its counterparts $A_\pm$ in
(\ref{neww4}) turn out to be self-adjoint operators in the Hilbert
spaces ${\mathfrak{L}_+}$ and ${\mathfrak{L}_-}$ with the inner
products ${[\cdot,\cdot]_{J}}$ and
$-{[\cdot,\cdot]_{J}}$, respectively. This simple
observation leads to the following statement.
\begin{proposition}[\cite{AKG}]\label{sese1}
A ${J}$-self-adjoint operator $A$ has the
property of ${C}$-symmetry if and only if $A$ is similar to a
self-adjoint operator in $\mathfrak{H}$.
If a $J$-self-adjoint operator $A$ has the property of $C$-symmetry then
its spectrum is real and the adjoint operator $C^*$ provides the
property of $C$-symmetry for $A^*$.
\end{proposition}

\begin{definition}\label{dad2}
Let ${A}\in\Sigma_J(S)$ have the property of $C$-symmetry realized by an operator $C$.  We will say that ${A}$
belongs to the sector $\Sigma_J^{\textsf{st}}$ of stable ${C}$-symmetry if
the operator $C$  commutes with $S$. Otherwise ($AC=CA$ but $SC\not=CS$), the operator
$A$ belongs to the sector $\Sigma_J^{\textsf{unst}}$ of unstable ${C}$-symmetry.
\end{definition}

The next statement immediately follows from Theorem 3.1 in
\cite{AKG}.

\begin{proposition}\label{ppp1}
Let $A_M\in\Sigma_J(S)$ be defined by (\ref{e55}).
Then $A_M\in\Sigma_J^{\textsf{st}}$ if and only if $CM=M$, where $C$ realizes the property
of $C$-symmetry for $S$.
\end{proposition}

\setcounter{equation}{0}
\section{The Phillips symmetric operator}
We are going to specify general results of previous section to the case of Phillips symmetric operator
$S$ defined by (\ref{e1}) and (\ref{e2}).
\subsection{Preliminaries.} The general definition (\ref{e1}), (\ref{e2}) of $S$
looks rather abstract and, in many cases, it is useful to work with
a model realization of $S$ in ${\mathfrak
H}=l_2(\mathbb{Z}, N)$ ($N$ is an auxiliary finite-dimensional
Hilbert space). In that case:
\begin{equation}\label{e3}
\begin{array}{l}
U(\ldots, x_{-2}, x_{-1}, \underline{x_{0}}, x_{1},
x_{2},\ldots)=(\ldots, x_{-3}, x_{-2}, \underline{x_{-1}}, x_{0},
x_{1},\ldots), \vspace{3mm} \\
V(\ldots, x_{-2}, x_{-1}, \underline{0}, x_{1},
x_{2},\ldots)=(\ldots, x_{-3}, x_{-2}, \underline{x_{-1}}, 0,
x_{1},\ldots),
\end{array}
\end{equation}
where $x_j\in{N}$ and elements at the zero position are underlined.

The self-adjoint operator $A$  takes the form:
\begin{equation}\label{e4}
\begin{array}{l}
Af=i(\ldots, x_{-3}+x_{-2}, x_{-2}+x_{-1}, \underline{x_{-1}+x_{0}},
x_{0}+x_{1}, x_{1}+x_{2}, \ldots) \vspace{3mm} \\
f\in\mathcal{D}(A)\iff{f}=(\ldots, x_{-3}-x_{-2}, x_{-2}-x_{-1},
\underline{x_{-1}-x_{0}}, x_{0}-x_{1}, x_{1}-x_{2}, \ldots),
\end{array}
\end{equation}
where $\sum_{i\in\mathbb{Z}}\|x_i\|_N^2<\infty$ and the symmetric operator $S$ is the restriction of
$A$ onto the set
\begin{equation}\label{ee5}
u\in\mathcal{D}(S)\iff{u}=(\ldots, x_{-3}-x_{-2},
x_{-2}-x_{-1}, \underline{x_{-1}},-x_{1}, x_{1}-x_{2}, \ldots),
\end{equation}
which consists of all $u\in\mathcal{D}(A)$ such that $x_0=0$.

Recalling (\ref{es15}) and using (\ref{e4}), (\ref{ee5}), it is easily to see that (see, e.g., \cite{KK})
\begin{equation}\label{sas3}
\begin{array}{l}
\mathfrak{N}_{i}=\{f_i(x)=(\ldots, 0, 0, \underline{x},
0, 0,\ldots) : \forall{x\in{N}}\}, \vspace{3mm} \\
\mathfrak{N}_{\mu}=\{f_\mu(x)=(\ldots, \overline{{r}}_{\mu}^2x,
\overline{r}_{{\mu}}x, \underline{x}, 0, 0 ,\ldots) :
\forall{x\in{N}}\}, \quad  \mu\in\mathbb{C}_+,  \vspace{3mm} \\
\mathfrak{N}_{-i}=\{f_{-i}(x)=(\ldots, 0, 0, \underline{0},
x, 0,\ldots) : \forall{x\in{N}}\}, \vspace{3mm} \\
\mathfrak{N}_{\overline{\mu}}=\{f_{\overline{\mu}}(x)=(\ldots, 0, 0, \underline{0}, x,
r_\mu{x}, r_\mu^2{x}, \ldots) : \forall{x\in{N}}\}.
\end{array}
\end{equation}
where $r_\mu=\frac{\mu-i}{\mu+i}$. Direct calculation with the use
of (\ref{ee5}) and (\ref{sas3}) gives
\begin{equation}\label{es1b}
f_\mu((1-\overline{r}_{{\mu}})x)=u+f_{i}(x), \quad
f_{\overline{\mu}}((1-{r}_{{\mu}})x)=v+f_{-i}(x), \quad
\forall{x}\in{N},
\end{equation}
where $u, v\in\mathcal{D}(S)$. Therefore,
\begin{equation}\label{es2}
\mathfrak{N}_{\mu}\subset\mathcal{D}(S)\dot{+}\mathfrak{N}_{i}, \quad \mathfrak{N}_{\overline{\mu}}\subset\mathcal{D}(S)\dot{+}\mathfrak{N}_{-i}, \quad  \forall\mu\in\mathbb{C}_+.
\end{equation}

\begin{lemma}\label{drdr}
Let $(\mathfrak{N}_{-i}, \Gamma_0,
\Gamma_1, Q)$ be a boundary triplet of the Phillips symmetric operator $S$ (defined by (\ref{e1}) and (\ref{e2})).
Then the corresponding characteristic function $\Theta(\cdot)$ of $S$ is equal to zero.
\end{lemma}

\emph{Proof.} It is sufficient to verify this statement for the case where $S$ is defined
by (\ref{e4}) and (\ref{ee5}). According to (\ref{es2}), an arbitrary $f_{\overline{\mu}}\in\mathfrak{N}_{\overline{\mu}}$
has the form $f_{\overline{\mu}}=u+f_{-i}$, where $u\in\mathcal{D}(S)$ and $f_{-i}\in\mathfrak{N}_{-i}$. But then
$(\Gamma_1+i\Gamma_0)f_{\overline{\mu}}=2if_{-i}$ and $(\Gamma_1-i\Gamma_0)f_{\overline{\mu}}=0$ due to (\ref{sas6}).
Therefore, $\Theta(\mu)\equiv{0}$ ($\forall{\mu}\in\mathbb{C}_+$) by (\ref{neww65}). Lemma \ref{drdr} is proved.

\begin{lemma}\label{lele1}
Let $S$ be defined by (\ref{e4}) and (\ref{ee5}) and let $J$ be a fundamental symmetry in $l_2(\mathbb{Z}, N)$.
Then $J$ commutes with $S$ if and only if
\begin{equation}\label{es14}
J(\ldots, x_{-2}, x_{-1}, \underline{x_{0}}, x_{1},
x_{2},\ldots)=(\ldots, J_{-}x_{-2}, J_{-}x_{-1},
\underline{J_{-}x_{0}}, J_+x_{1}, J_+x_{2},\ldots),
\end{equation}
where $J_{\pm}$  are fundamental symmetries in $N$.
\end{lemma}

{\it Proof.} Let $J$ commute with $S$. It follows from (\ref{es15}) that defect subspaces $\mathfrak{N}_\mu$ are invariant with respect $J$.
Taking (\ref{sas3}) into account we conclude that the restrictions $J_-:=J\upharpoonright\mathfrak{N}_{i}$ and $J_+:=J\upharpoonright\mathfrak{N}_{-i}$ determine two fundamental symmetries $J_-$ and $J_+$ in $N$.
Further, the equality $JS=SJ$ is equivalent to the relation
$JV=VJ$, where $V$ is defined by (\ref{e3}).
Combining this relation with the first and third relations in (\ref{sas3}) and taking the definition of $J_\pm$ into account
we establish (\ref{es14})

Conversely, if a fundamental symmetry $J$ is defined by (\ref{es14}), then relations (\ref{e4}) and (\ref{ee5}) imply that $JS=SJ$.
Lemma \ref{lele1} is proved.

\begin{lemma}\label{lele2}
Let $S$ be defined by (\ref{e4}) and (\ref{ee5}), let $J$ be a fundamental symmetry in $l_2(\mathbb{Z}, N)$ commuting with $S$,
and let $C$ be a bounded operator in $l_2(\mathbb{Z}, N)$ such that $C^2=I$ and $JC>0$. Then $C$ commutes with $S$ if and only if
\begin{equation}\label{es14c}
C(\ldots, x_{-2}, x_{-1}, \underline{x_{0}}, x_{1},
x_{2},\ldots)=(\ldots, C_{-}x_{-2}, C_{-}x_{-1},
\underline{C_{-}x_{0}}, C_+x_{1}, C_+x_{2},\ldots),
\end{equation}
where $C_{\pm}$ are bounded
operators in $N$ such that $C_{\pm}^2=I_N$ and $J_{\pm}C_{\pm}>0$ where $J_{\pm}$ are taken from the formula (\ref{es14}).
\end{lemma}

{\it Proof.} By Lemma \ref{lele1}, the operator $J$ is defined by (\ref{es14}), where
$J_{\pm}$  are fundamental symmetries in $N$.

Assume that $C$ commutes with $S$. Then, using (\ref{es12})
one gets $SF=FS$, where $F=JC$ is a bounded
self-adjoint operator. Hence,
$$
S{C}^{*}=SFJ=FSJ=FJS={C}^{*}S.
$$
The obtained relation
${C}^{*}S=S{C}^{*}$
and ${C}^2=I$ imply that the defect subspaces $\mathfrak{N}_\mu$ of $S$ are invariant with respect $C$.
It follows from (\ref{sas3}) that the restrictions $C_-:=C\upharpoonright\mathfrak{N}_{i}$ and $C_+:=C\upharpoonright\mathfrak{N}_{-i}$ determine bounded operators $C_{\pm}$ in $N$ such that $C_{\pm}^2=I_N$ and $J_{\pm}C_{\pm}>0$.
Reasoning by analogy with the proof of Lemma \ref{lele1}, we complete the proof.

\subsection{Description of $J$-self-adjoint extensions.}
Using (\ref{sas3}) we can identify the Hilbert space $\mathfrak{M}=\mathfrak{N}_{-i}\dot{+}\mathfrak{N}_{i}$ with
$$
N\oplus{N}=\left\{\left(\begin{array}{c}
x  \\
y
\end{array}\right) \ | \ x, y\in{N}\right\}.
$$
In that case
\begin{equation}\label{es5}
Z=\left(\begin{array}{cc}
I & 0 \\
0 & -I
\end{array}\right) \quad \mbox{and} \quad J\upharpoonright\mathfrak{M}=\left(\begin{array}{cc}
J_+ & 0 \\
0 & J_-
\end{array}\right).
\end{equation}
\begin{proposition}\label{p1234}
Let $S$ be defined by (\ref{e4}) and (\ref{ee5}). Then the set $\Sigma_J(S)$ of $J$-self-adjoint extensions of $S$ is non-empty if and only if
\begin{equation}\label{es6}
\dim[(I-J_+)N]=\dim[(I-J_-)N].
\end{equation}
\end{proposition}

{\it Proof.} By Proposition \ref{t2}, $J$-self-adjoint extensions of $S$ exist
if and only if the Krein space $(\mathfrak{M}, [\cdot, \cdot]_{JZ})$ has hypermaximal neutral subspaces.
This is possible only in the case where $\dim[(I+JZ)\mathfrak{M}]=\dim[(I-JZ)\mathfrak{M}]$ or, that is equivalent (see (\ref{es5})),
$$
\dim[(I+J_+)N]+\dim[(I-J_-)N]=\dim[(I-J_+)N]+\dim[(I+J_-)N].
$$
This identity is equivalent to (\ref{es6}) (since $\dim[(I+J_\pm)N]+\dim[(I-J_\pm)N]=\dim{N}$ and $\dim{N}<\infty$).
Proposition \ref{es6} is proved.

\begin{corollary}\label{cc1}
Let $S$ be defined by (\ref{e4}) and (\ref{ee5}) and let $J$ be a fundamental symmetry commuting with $S$ in $l_2(\mathbb{Z}, N)$.
Then self-adjoint extensions of $S$ commuting with $J$ exist if and only if
the identity (\ref{es6}) holds.
\end{corollary}

{\it Proof.} If $A_M$ is a self-adjoint extension of  $S$ commuting with $J$, then $A_M\in\Sigma_J(S)$
 and relation (\ref{es6}) holds due to Proposition \ref{p1234}.

Conversely, since $\dim{N}<\infty$, relation (\ref{es6}) is
equivalent to the identity
$$
\dim[(I+J_+)N]=\dim[(I+J_-)N].
$$
This implies the existence of unitary mappings $G
: \mathfrak{N}_{i}\to\mathfrak{N}_{-i}$ such that $GJ=GJ_-=J_+G=JG$.
In that case the hypermaximal neutral subspace $M_G$ of the Krein
space $(\mathfrak{M}, [\cdot, \cdot]_{Z})$ (defined by (\ref{sas14}))
satisfies the relation $JM_G=M_G$ and the corresponding self-adjoint
extension $A_M$ commutes with $J$. Corollary \ref{cc1} is proved.

\begin{proposition}\label{cc2a}
Let $S$ be the Phillips symmetric operator (defined by (\ref{e1}) and (\ref{e2})) and
let $J$ be a fundamental symmetry commuting with $S$ in $\mathfrak{H}$. Then
boundary triplets $(\mathfrak{N}_{-i}, \Gamma_0, \Gamma_1, Q)$ of $S^*$ defined by (\ref{sas6}) and satisfying
(\ref{ea6a}) exist if and only if the set $\Sigma_J(S)$ is non-empty.
\end{proposition}

\emph{Proof.} It is sufficient to establish for the Phillips symmetric operator $S$ realized by the formulas
(\ref{e4}) and (\ref{ee5}). In that case, by Proposition \ref{p1234} and Corollary \ref{cc1},
 $\Sigma_J(S)\not=\emptyset$ $\iff$ there exist
self-adjoint extensions of $S$ commuting with $J$.
Using now Proposition \ref{cc2} we complete the proof.

\begin{theorem}\label{tt1}
Let $S$ be the Phillips symmetric operator,
let $J$ be a fundamental symmetry commuting with $S$ in $\mathfrak{H}$, and let $A_M\in\Sigma_J(S)$.
Then the spectrum of $A_M$ either coincides with $\mathbb{R}$ ($\sigma(A_M)=\mathbb{R}$) or covers the whole complex
plane ($\sigma(A_M)=\mathbb{C}$) and its non-real part consists of eigenvalues of $A_M$.
\end{theorem}

{\it Proof.}  Since an arbitrary $A_M\in\Sigma_J(S)$ is a finite rank perturbation of the self-adjoint
operator $A$ (see (\ref{e2})), the non-real spectrum of $A_M$ may include complex eigenvalues.

Without loss of generality we can suppose that $S$ is determined by the formulas
(\ref{e4}) and (\ref{ee5}). Assume that $\mu_0\in\mathbb{C}_+$ is an eigenvalue
of $A_M$. Then there exists an element
$f_{\overline{\mu}_0}\in\mathfrak{N}_{\overline{\mu}_0}\cap\mathcal{D}(A_M)$ and,
according to (\ref{es1b}),
$$
f_{\overline{\mu}_0}=f_{\overline{\mu}_0}(x)=v_0+f_{-i}\left(\frac{x}{1-{r}_{{\mu_0}}}\right), \quad v_0\in\mathcal{D}(S)
$$
for a certain choice of $x\in{N}$. Using the second relation in (\ref{es1b}) for an arbitrary $\mu\in\mathbb{C}_+$, we obtain
$$
f_{\overline{\mu}}\left(\frac{1-{r}_{{\mu}}}{1-{r}_{{\mu_0}}}x\right)=v+f_{-i}\left(\frac{x}{1-{r}_{{\mu_0}}}\right), \quad v\in\mathcal{D}(S)
$$
Comparing last two relations we arrive at the conclusion that the element
$$
f_{\overline{\mu}}\left(\frac{1-{r}_{{\mu}}}{1-{r}_{{\mu_0}}}x\right)=v-v_0+f_{\overline{\mu}_0}, \qquad \mu\in\mathbb{C}_+.
$$
belongs to $\mathfrak{N}_{\overline{\mu}}\cap\mathcal{D}(A_M)$.  
Hence $\mathbb{C}_+\subset\sigma_p(A_M)$. The relation
$\mathbb{C}_-\subset\sigma_p(A_M)$ is established by the same manner. Thus,
$\sigma(A_M)=\mathbb{C}$ and $\mathbb{C}\setminus\mathbb{R}$
contains eigenvalues of $A_M$.

Assume that the spectrum of $A_M$ is real. Since the Phillips symmetric operator has no real points of regular type
(see, e.g., \cite{KO}), the spectrum of $A_M$ coincides with $\mathbb{R}$.
Theorem \ref{tt1} is proved.

\begin{corollary}\label{k12c}
Let $S$ be the Phillips symmetric operator and
let $J$ be a fundamental symmetry commuting with $S$ in $\mathfrak{H}$.
Then the set $\Sigma_J(S)$ does not contain definitizable operators.
\end{corollary}

\emph{Proof.} By Proposition (\ref{cc2a}), if $\Sigma_J(S)$ is a non-empty set, then there exists a boundary triplet
 $(\mathfrak{N}_{-i}, \Gamma_0, \Gamma_1, Q)$ of $S^*$ which satisfies (\ref{ea6a}).
It follows from Proposition \ref{tt34} and Theorem \ref{tt1} that operators $A_M\in\Sigma_J(S)$ with \emph{real} spectrum
are described by the formula (\ref{as1}) in terms of the boundary triplet $(\mathfrak{N}_{-i}, \Gamma_0, \Gamma_1, Q)$. The rest of
 operators $A_M\in\Sigma_J(S)$ (which can not be described by (\ref{as1})) have empty resolvent set (due to Remark \ref{rr1} and Theorem \ref{tt1}). By Proposition \ref{te2e} this means that $\Sigma_J(S)$ does not contain definitizable operators.
Corollary \ref{k12c} is proved.

\subsection{$J$-self-adjoint extensions with ${C}$-symmetry.}
\begin{theorem}\label{p14a}
Let $S$ be the Phillips operator with deficiency indices $<2,2>$. Then an arbitrary $J$-self-adjoint extension  $A_M\in\Sigma_J(S)$ has the property of stable $C$-symmetry ($A_M\in\Sigma_J^{\textsf{st}}$) if and only if the spectrum of $A_M$ is real.
\end{theorem}
\emph{Proof.}
If $A_M$ has $C$-symmetry, then its spectrum is real (see Proposition \ref{sese1}).

Conversely, we assume that
$A_M\in\Sigma_J(S)$ has a real spectrum.
In that case, by Proposition \ref{k12}, $A_M(=A_{\mathcal{K}})$ is defined by (\ref{as1b}), 
where $\mathcal{K}=\mathcal{K}(\zeta, \phi, \omega, \xi)$ has the form (\ref{as3a}).

Without loss of generality we can assume that $S$ is determined by the formulas
(\ref{e4}) and (\ref{ee5}) in the space $l_2(\mathbb{Z}, N)$. Then, 
by virtue of Proposition \ref{ppp1} and Lemma \ref{lele2}, the operator
$A_M(=A_{\mathcal{K}})$ has a stable $C$-symmetry if and only if $CM=M$ for at least one
of operators $C$ determined by (\ref{es14c}). 

It follows from (\ref{as1b}) that $M=\{f=f_{-i}+f_i \ | \
K(\Gamma_1+i\Gamma_0)f=(\Gamma_1-i\Gamma_0)f\}$. Employing
(\ref{sas6}) we rewrite the latter relation as follows
$$
M=\{f_{-i}-Q^{-1}Kf_{-i} \ | \ \forall{f_{-i}}\in\mathfrak{N}_{-i}\}.
$$
The obtained description of $M$ and Lemma \ref{lele2} imply that
\begin{equation}\label{new29}
CM=M \qquad \iff \qquad KC_+=\widehat{C}_+K  \quad
(\widehat{C}_+:=QC_-Q^{-1}),
\end{equation}
where $C_+$ and $\widehat{C}_+$ act in $N=\mathfrak{N}_{-i}$ and satisfy the relations
$C^2=I$, \ $JC>0$ \ $(C\in\{C_+, \widehat{C}_+\})$.

Let ${\mathcal{C}}_+=\|c_{ij}\|$ be the matrix representation of ${C}_+$ with respect to
the basis $\{e_{++},e_{+-}\}$. Then the relations $C_+^2=I$, \ $JC_+>0$ take the form
\begin{equation}\label{sssr4}
{\mathcal{C}}_+^2=\left(\begin{array}{cc}
1 & 0 \\
0 & 1 \end{array}\right), \qquad  \left(\begin{array}{cc}
1 & 0 \\
0 & -1 \end{array}\right){\mathcal{C}}_+>0,
\end{equation}
 where  $\left(\begin{array}{cc}
1 & 0 \\
0 & -1 \end{array}\right)$ is the matrix representation of $J\upharpoonright\mathfrak{N}_{-i}$ with respect
to $\{e_{++},e_{+-}\}$ (since (\ref{ess1})). A simple analysis of (\ref{sssr4}) leads to the following description of
${\mathcal{C}}_+$:
\begin{equation}\label{sas8}
{\mathcal{C}}_+={\mathcal{C}}_{\widetilde{\chi},\widetilde{\omega}}
=\left(\begin{array}{cc}
\cosh\widetilde{\chi} & (\sinh\widetilde{\chi})e^{-i\widetilde{\omega}} \\
-(\sinh\widetilde{\chi})e^{i\widetilde{\omega}} & -\cosh\widetilde{\chi}
\end{array}\right), \qquad \widetilde{\chi}, \widetilde{\omega}\in\mathbb{R}.
\end{equation}

Reasoning by analogy for the matrix representation $\widehat{\mathcal{C}}_+$ of $\widehat{C}_+$ we get
\begin{equation}\label{sas8b}
\widehat{\mathcal{C}}_+={\mathcal{C}}_{\widehat{\chi},\widehat{\omega}}
=\left(\begin{array}{cc}
\cosh\widehat{\chi} & (\sinh\widehat{\chi})e^{-i\widehat{\omega}} \\
-(\sinh\widehat{\chi})e^{i\widehat{\omega}} & -\cosh\widehat{\chi}
\end{array}\right), \qquad \widehat{\chi}, \widehat{\omega}\in\mathbb{R}.
\end{equation}

Passing to the matrix representation in (\ref{new29})
we conclude that $A_M(=A_{\mathcal{K}})$ has a stable $C$-symmetry if and only if
\begin{equation}\label{sssr3}
\mathcal{K}(\zeta,
\phi, \omega, \xi){\mathcal{C}}_{\widetilde{\chi},\widetilde{\omega}}={\mathcal{C}}_{\widehat{\chi},\widehat{\omega}}\mathcal{K}(\zeta,
\phi, \omega, \xi)
\end{equation}
where $\mathcal{K}(\zeta,
\phi, \omega, \xi)$ is defined by (\ref{as3a}). A routine analysis of (\ref{sssr3}) with the use of (\ref{sas8}) and (\ref{sas8b}) shows that
(\ref{sssr3}) is equivalent to the system of relations
\begin{equation}\label{sssr14}
\left\{\begin{array}{l}
\cosh\widehat{\chi}-\cosh\widetilde{\chi}+\tanh\zeta[e^{i(\widehat{\omega}-\omega-\phi)}\sinh\widehat{\chi}-e^{i({\omega}-\widetilde{\omega}-\phi)}\sinh\widetilde{\chi}]=0 \vspace{3mm} \\
\tanh\zeta[\cosh\widehat{\chi}+\cosh\widetilde{\chi}]+e^{i(\phi+\omega-\widehat{\omega})}\sinh\widehat{\chi}+e^{-i(\phi-\omega+\widetilde{\omega})}\sinh\widetilde{\chi}=0.
\end{array}
\right.
\end{equation}

Let us set $\widehat{\chi}=\widetilde{\chi}=\chi$. Then the first relation in (\ref{sssr14}) is satisfied when
\begin{equation}\label{led2}
\omega=\frac{\widehat{\omega}+\widetilde{\omega}}{2}
\end{equation}
and the second one goes over
$$
\tanh\zeta+\tanh\chi\cos\left(\phi+\frac{\widetilde{\omega}-\widehat{\omega}}{2}\right)=0.
$$
The latter equation can be solved with respect to $\chi$ if and only if
\begin{equation}\label{led1}
|\tanh\zeta|<\left|\cos\left(\phi+\frac{\widetilde{\omega}-\widehat{\omega}}{2}\right)\right|.
\end{equation}

Since $\widetilde{\omega}, \widehat{\omega}\in\mathbb{R}$ are independent variables, conditions (\ref{led2}) and (\ref{led1}) with fixed
$\omega$ and $\phi$ can easily be satisfied by a suitable choice of  $\widetilde{\omega}$ and $\widehat{\omega}$.
This means that the system (\ref{sssr14}) has a solution $\widetilde{\chi}, \widehat{\chi}, \widetilde{\omega}, \widehat{\omega}$ for any fixed $\zeta,
\phi, \omega, \xi$. Therefore, $A_{\mathcal{K}(\zeta,
\phi, \omega, \xi)}$ has a stable $C$-symmetry for any choice of $\zeta,
\phi, \omega$, and  $\xi$.
Theorem \ref{p14a} is proved.

\begin{corollary}\label{uma1}
Let ${A}$ be $J$-self-adjoint extension of the Phillips symmetric
operator $S$ with deficiency indices $<2,2>$. Then ${A}$ is
similar to a self-adjoint operator if and only if the spectrum of
${A}$ is real.
\end{corollary}

{\it Proof.} It follows from Proposition \ref{sese1} and Theorem
\ref{p14a}

\subsection{Various realizations of the Phillips operator.}
It follows from (\ref{e1}) and (\ref{e2}) that the Phillips symmetric operator $S$ can be obtained as the restriction of a
self-adjoint operator $A$ with Lebesgue spectrum onto the domain
\begin{equation}\label{new19}
\mathcal{D}(S)=\{f\in\mathcal{D}(A) \ |  \ ((A-iI)f, w)=0, \ \forall{w}\in{W_0}\},
\end{equation}
where $W_0$ is a wandering subspace of the bilateral shift $U$. In particular, this means that the Phillips
symmetric operator $S$ naturally arises in the study of the formal expression $i\frac{d}{dx}+<\delta,\cdot>\delta(x)$ and it coincides with the
operator
$$
S=i\frac{d}{dx}, \quad \mathcal{D}(S)=\{u(\cdot)\in{W_2^1}(\mathbb{R}) \ | \ u(0)=0\}
$$
acting in $L_2(\mathbb{R})$.  This example illustrates one of possible general approaches to the construction
of the Phillips symmetric operator. Indeed, let $\mathfrak{H}={\mathfrak H}_1\oplus{\mathfrak H}_2$ and let
 $S=S_1\oplus{S_2}$, where $S_1$ and $S_2$ are simple maximal symmetric operators in the Hilbert spaces ${\mathfrak H}_1$ and ${\mathfrak H}_2$ with
deficiency indices $<m, 0>$ and $<0, m>$ ($m\in\mathbb{N}$), respectively. In that case $S$ is a simple symmetric operator
in $\mathfrak{H}$ with deficiency indices $<m, m>$ and its characteristic function $\Theta(\cdot)$ associated with an arbitrary
boundary triplet $(\mathfrak{N}_{-i}, \Gamma_0,
\Gamma_1, Q)$ (see (\ref{sas6})) is equal to zero. By Lemma \ref{drdr} this means that $S$ is a Phillips symmetric operator.

Using (\ref{e2}) and (\ref{new19}) it is easy to calculate the defect subspaces $\mathfrak{N}_{\pm{i}}$ of  $S$:
$$
\mathfrak{N}_{{i}}=W_0 \quad \mbox{and} \quad \mathfrak{N}_{{-i}}=UW_0.
$$

According to (\ref{e1}) and (\ref{e2}), a bilateral shift $U$ and its wandering subspace $W_0$ are the main ingredients for the determination of
the Phillips symmetric operator $S$. To illustrate this point, we have presented below two mathematical constructions where $U$ appears
naturally and $W_0$ admits a simple description.

\subsubsection{Multiresolution approximation of $L_2(\mathbb{R})$.}

We recall \cite{MAL,ME} that a multiresolution approximation (MRA) of $L_2(\mathbb{R})$ is a sequence
$\{V_j\}_{j\in\mathbb{Z}}$ of closed subspaces of $L_2(\mathbb{R})$ such that:
$(i) \ V_j\subset V_{j+1}, \ j\in\mathbb{Z}$; \quad
$(ii) \ \cap_{j\in\mathbb{Z}}V_j =\{0\}$; \quad $(iii)\ \cup_{j\in\mathbb{Z}}V_j$ is dense in $L_2(\mathbb{R})$;
\quad $(iv) \ f(\cdot)\in{V}_j\Leftrightarrow{f}(2^{-j}\cdot)\in{V}_0$; \quad  $(v)$ \ there exists a function $\varphi(\cdot)\in{V}_0$ such that the sequence $\{\varphi(\cdot-k), k\in\mathbb{Z}\}$ is a Riesz basis of $V_0$.

Let
$$
Uf(x)=\frac{1}{\sqrt{2}}f\left(\frac{x}{2}\right), \qquad \forall{f}\in{L_2(\mathbb{R})}
$$
be the dilation operator in $\mathfrak{H}={L_2(\mathbb{R})}$ and let $\{V_j\}_{j\in\mathbb{Z}}$ be a fixed multiresolution approximation of $L_2(\mathbb{R})$. Then $U$ is a bilateral shift in $L_2(\mathbb{R})$ with a wandering subspace $W_0=V_1\ominus{V_0}$ (due to properties $(i)-(iv)$).

According to the general results of MRA-based wavelet theory \cite{MAL,ME} the subspace $W_0$ is a wavelet subspace and relations $(i)-(v)$ imply the existence of a function (wavelet) $\psi(\cdot)\in{W_0}$ such that the sequence $\{\psi(\cdot-k), k\in\mathbb{Z}\}$ forms an orthonormal basis in $W_0$. This means that (\ref{new19}) can be rewritten as follows:
$$
\mathcal{D}(S)=\{f\in\mathcal{D}(A) \ |  \ ((A-iI)f, \psi(\cdot-k))=0, \ \forall{k}\in\mathbb{Z}\},
$$
where the wavelet $\psi(\cdot)$ is directly constructed by the (scaling) function $\varphi(\cdot)$ from condition $(v)$.

\subsubsection{Abstract wave equation.}
Let us consider an operator-differential equation
\begin{equation}\label{es1a}
  u_{tt}=-Lu,
\end{equation}
where $L$ is a positive self-adjoint operator in a Hilbert
space  $H$. By ${H}_L$ we denote the
completion of domain of definition $\mathcal{D}(L)$ with respect to the
norm $\|u\|_{{H}_L}^2:=(Lu,u)_{H}$ and consider the Hilbert space $\mathfrak{H}={H}_L\oplus{H}$ (the energy space).
It is convenient to write elements of $\mathfrak{H}$ as
column matrices $\left(\begin{array}{c}
 u \\
 v \end{array}\right)$, where $u\in{H}_L$ and $v\in{H}$.
 Put $u_t=v$ and rewrite (\ref{es1a}) as
 $$ {\frac{d}{dt}}\left(\begin{array}{cc}
 u \\  v \end{array} \right)=iQ\left(\begin{array}{cc}
 u \\  v \end{array} \right), \qquad
 Q=i\left(\begin{array}{cc}
 0 & -I \\
 L & 0 \end{array}\right).
 $$
 The operator $Q$ with the domain
 $
 \mathcal{D}(Q)=\left\{\left(\begin{array}{c}
 u \\
 v \end{array}\right) \ | \ \{u, v\}\subset{\mathcal{D}(L)}\right\}$
 is essentially self-adjoint in $\mathfrak{H}$. Its closure
 $A$ is a generator of the group of unitary (in $\mathfrak{H}$)
 operators $W_{A}(t)=e^{iAt}$, which determines solutions of the Cauchy problem
 for the abstract wave equation (\ref{es1a}).

The equation (\ref{es1a}) is said to be \emph{free (unperturbed) wave equation} if there exists a simple maximal symmetric operator $B$ in ${H}$ such that
\begin{equation}\label{es2a}
B^2\subset{L} \qquad \mbox{and} \qquad (Lu,u)=\|B^*u\|_H^2, \quad \forall{u}\in\mathcal{D}(L).
\end{equation}

Assume that $\{u_n\}$ belong $\mathcal{D}(B^2)$ and form a Cauchy sequence in $H_L$. Then $\{Bu_n\}$ is the Cauchy sequence in $H$
(due to the second relation in (\ref{es2a})) and hence
$\lim_{n\to\infty}Bu_n=\gamma\in{H}$. In that case we will say that the sequence $\{u_n\}$ converges to the element $\textsf{x}_\gamma$ in the space $H_L$. Obviously the Hilbert space $H_L$ can be identified with the set of elements $\{\textsf{x}_\gamma \ | \ \forall\gamma\in{H}\}$ and
$(\textsf{x}_\gamma, \textsf{x}_\zeta)_{H_L}=(\gamma, \zeta)_H$.

In what follows, without loss of generality we assume that $B$ has zero defect number in the lower half-plane. Then $B$ admits
the representation
\begin{equation}\label{e67}
B=T^{-1}i\frac{d}{ds}T, \qquad  \mathcal{D}(B)=T^{-1}\stackrel{0}{W^1_2}({\Bbb R}_+,\mathcal{N}),
\end{equation}
where $T$ is an isometric mapping from ${H}$ onto $L_2({\mathbb{R}}_+,\mathcal{N})$,
$\mathcal{N}$ is an auxiliary Hilbert space of dimension equal to the nonzero defect number of $B$ and ${\mathbb{R}}_+=(0,\infty)$.

\par
Using (\ref{e67}), we can define $B$ in various functional spaces getting, as a result, different  specific realizations of the
free abstract wave equation. In particular, the classical free wave equation $u_{tt}(x,t)=\Delta{u}(x,t)$ in $\mathbb{R}^n$ ($n$ is odd)
can be obtained from (\ref{e67}) if we choose $\mathcal{N}$ as the Hilbert space $L_2(S^{n-1})$ of functions square-integrable on the unit
sphere $S^{n-1}$ in ${\mathbb{R}}^n$ and consider the isometric operator $T :L_2({\mathbb{R}}^n)\to{L_2({\mathbb{R}}_+,\mathcal{N})}$
defined on the rapidly decreasing smooth functions $u(x)\in{S({\Bbb R}^n)}$ by the formula
$$
(Tu)(s,w)=(\partial_s^mRu)(s,w) \qquad m=\frac{(n-1)}{2}, \quad  s\ge0, \quad w\in{S^{n-1}},
$$
where $R$ is the Radon transformation. In that case the Laplace operator $L=-\Delta$ in $L_2({\mathbb{R}}^n)$ satisfies
condition (\ref{es2a}) and equation (\ref{es1a}) takes the form $u_{tt}(x,t)=\Delta{u}(x,t)$ (see \cite{KU2} for detail).

Assume that (\ref{es1a}) is the free wave equation for some choice of $B$. Then the corresponding generator $A$ is the
Cayley transform of a bilateral shift $U$ and a wandering subspace $W_0$ can be chosen as follows \cite{KU11}:
\begin{equation}\label{new199}
W_0=\left\{\left(\begin{array}{c}
\textsf{x}_h \\
-ih
\end{array}\right) \ | \ \forall{h}\in\ker(B^*+iI) \right\}.
\end{equation}
Substituting (\ref{new199}) into (\ref{new19}) and taking into account that the domain $\mathcal{D}(A)$ can be described explicitly,
we find $S$. For instance, let $L$ be the Friedrichs extension of $B^2$. Then $L=B^*B$ and this operator satisfies
(\ref{es2a}).  In this case:
$$
A\left(\begin{array}{c}
\textsf{x}_\gamma \\
p \end{array}\right)=-i\left(\begin{array}{c}
\textsf{x}_{Bp} \\
-B^*\gamma \end{array}\right), \quad \mathcal{D}(A)=\left\{\left(\begin{array}{c}
\textsf{x}_\gamma \\
p \end{array}\right) \ | \ \forall{\gamma}\in\mathcal{D}(B^*), \ \forall{p}\in\mathcal{D}(B) \right\}
$$
and $S$ is the restriction of $A$ onto the set of elements
$$
\left\{\left(\begin{array}{c}
\textsf{x}_\gamma \\
p \end{array}\right) \quad | \quad  {\gamma}\in\mathcal{D}(B^*), \  {p}\in\mathcal{D}(B)\right\}
$$
such that $((B^*+iI)(p-i\gamma), h)=0$ for all ${h}\in\ker(B^*+iI)$.
\begin{corollary}
Assume that the nonzero defect number of $B$ is $2$ and $J$ is a fundamental symmetry in $\mathfrak{H}$ such that $SJ=JS$. Then if ${A}\in\Sigma_J(S)$ has a real spectrum, then
$W_{{A}}(t)=e^{i{A}t}$ is a $C_0$-semigroup.
\end{corollary}

\emph{Proof.} Immediately follows from Corollary \ref{uma1}.

\section{Conclusions.} In this paper we have studied the collection $\Sigma_J(S)$ of  $J$-self-adjoint extensions of the Phillips symmetric operator $S$. Our attention to $\Sigma_J(S)$ was inspirited by a steady interest in the spectral analysis of new classes of $J$-self-adjoint operators $A_\varepsilon$ with the aim to illustrate quantitative and qualitative changes of spectra
$\sigma(A_\varepsilon)$ when parameters $\varepsilon$ run the domain of variation $\Xi$.
Due to specific inherent properties of the Phillips operator $S$ (the zero characteristic function, the absence of real points of regular type, etc) we obtained a spectral picture which differs from the matrix models \cite{GU1,GRS,GSZ} and models based on $J$-self-adjoint
(symmetric) perturbations of the Schr\"{o}dinger or the Dirac operator \cite{AMS,CGS,LT,Z}. For instance, in our case, either the spectrum of  $A\in\Sigma_J(S)$ coincides with real line: $\sigma(A)=\mathbb{R}$ or with complex plane: $\sigma(A)=\mathbb{C}$ (Theorem \ref{tt1}).

For operators $A_\varepsilon\in\Sigma_J(S)$ (where $S$ is an arbitrary symmetric operator commuting with $J$) we have introduced the concepts of stable and unstable $C$-symmetry (Definition \ref{dad2}). These concepts are natural
for sets of $J$-self-adjoint operators appearing in the extension theory framework. Roughly speaking, if $A_\varepsilon$ belongs to the sector $\Sigma_J^{\textsf{st}}$ of stable $C$-symmetry, then $A_\varepsilon$ preserves the property of $C$-symmetry under small variation
of $\varepsilon$.

For singular perturbations of the Schr\"{o}dinger or the Dirac operator, the corresponding symmetric operator
$S$ has real points of regular type. In that case, the sector $\Sigma_J^{\textsf{unst}}$ of unstable $C$ symmetry
is not empty and operators $A_\varepsilon\in\Sigma_J(S)$ with real spectra and Jordan points arise in the case
where $\varepsilon$ lies on the boundary of $\Sigma_J^{\textsf{st}}$ \cite{AKG, GK1}.
This picture is essentially simplified for the Phillips symmetric operator $S$ since $S$ has no
real points of regular type. We have shown that the sector $\Sigma_J^{\textsf{unst}}$ of unstable ${C}$-symmetry is the empty set
and there are no $J$-self-adjoint extensions of $A_{\mathrm{sym}}$ with real spectra and Jordan points (this fact follows from Theorem \ref{p14a} and Corollary \ref{uma1}). These results have been obtained under the assumption that $S$ has deficiency indices $<2,2>$.
We believe that they remain true for the general case $<n,n>$. However, the corresponding proof requires more cumbersome analysis
and the case $<n,n>$ will be considered in a forthcoming paper.

An open problem is finding an adequate physical phenomenon for which $J$-self-adjoint extensions
of $S$ can be served as model Hamiltonians. In this way we have just discussed certain representations of $S$
related to abstract wave equation and multiresolution approximation.

\bigskip\noindent \textbf{Acknowledgements.} The first author (S.K) thanks Yu.~Arlinskii for useful discussions which led to improvements 
in the paper and DFFD of Ukraine (research projects F28.1/017 and F29.1/010) for the support.

\end{document}